\documentclass[fleqn,usenatbib]{mnras}

\usepackage{newtxtext,newtxmath}

\usepackage[T1]{fontenc}

\DeclareRobustCommand{\VAN}[3]{#2}
\let\VANthebibliography\thebibliography
\def\thebibliography{\DeclareRobustCommand{\VAN}[3]{##3}\VANthebibliography}

\usepackage{graphicx}	
\usepackage{amsmath}	

\title[MUSE-ALMA Haloes X: stellar masses determination]{MUSE-ALMA Haloes X: The stellar masses of gas-rich absorbing galaxies}

\author[R. Augustin et al.]{
Ramona Augustin,$^{1,2}$\thanks{E-mail: raugustin@aip.de}
C\'eline P\'eroux,$^{3,4}$
Arjun Karki,$^{5}$
Varsha Kulkarni,$^{5}$
Simon Weng,$^{3,6,7,8}$
\newauthor
A. Hamanowicz,$^{1}$
M. Hayes,$^{9}$ 
J. C. Howk,$^{10}$
G. G. Kacprzak,$^{11, 7}$
A. Klitsch,$^{12}$
M. A. Zwaan,$^{3}$
A. Fox,$^{13,14}$
\newauthor
A. Biggs,$^{3}$
A. Y. Fresco,$^{15}$ 
S. Kassin,$^{1}$
H. Kuntschner$^{3}$\\
$^{1}$ Space Telescope Science Institute, 3700 San Martin Drive, Baltimore, MD 21218, USA\\
$^{2}$ Leibniz-Institut f{\"u}r Astrophysik Potsdam (AIP), An der Sternwarte 16, 14482 Potsdam, Germany\\
$^{3}$ European Southern Observatory (ESO), Karl-Schwarzschild-Str. 2, 85748 Garching bei M{\"u}nchen, Germany\\
$^{4}$ Aix Marseille Universit\'e, CNRS, LAM (Laboratoire d'Astrophysique de Marseille) UMR 7326, 13388, Marseille, France\\
$^5$ Department of Physics and Astronomy, University of South Carolina, Columbia, SC 29208, USA\\
$^6$ Sydney Institute for Astronomy, School of Physics A28, University of Sydney, NSW 2006, Australia\\
$^7$ ARC Centre of Excellence for All Sky Astrophysics in 3 Dimensions (ASTRO 3D)\\ 
$^8$ ATNF, CSIRO Space and Astronomy,  PO Box 76, Epping, NSW 1710, Australia\\
$^{9}$ Stockholm University, Department of Astronomy and Oskar Klein Centre for Cosmoparticle Physics, AlbaNova University Centre, SE-10691, Stockholm, Sweden\\
$^{10}$ Department of Physics, University of Notre Dame, Notre Dame, Indiana 46556, USA\\
$^{11}$ Centre for Astrophysics and Supercomputing, Swinburne University of Technology, Hawthorn, Victoria 3122, Australia\\
$^{12}$ DARK, Niels Bohr Institute, University of Copenhagen, Jagtvej 128, 2200 Copenhagen, Denmark\\
$^{13}$AURA for ESA, Space Telescope Science Institute, 3700 San Martin Drive, Baltimore, MD 21218\\
$^{14}$ Department of Physics \& Astronomy, Johns Hopkins University, 3400 N. Charles Street, Baltimore, MD 21218, USA\\
$^{15}$ Max-Planck-Institut f{\"u}r Extraterrestrische Physik (MPE), Giessenbachstrasse 1, D--85748 Garching, Germany\\
}

\date{Accepted XXX. Received YYY; in original form ZZZ}

\pubyear{2015}

\begin{document}
\label{firstpage}
\pagerange{\pageref{firstpage}--\pageref{lastpage}}
\maketitle

\begin{abstract}
The physical processes by which gas is accreted onto galaxies, transformed into stars and then expelled from
galaxies are of paramount importance to galaxy evolution studies. Observationally constraining each of these
baryonic components in the same systems however, is challenging. 
Furthermore, simulations indicate that the stellar mass of galaxies is a key factor influencing CGM properties.
Indeed, absorption lines detected against background quasars offer the most compelling way to study the cold gas in the circumgalactic medium (CGM).
The MUSE-ALMA Haloes survey is composed of quasar fields covered with VLT/MUSE observations, comprising 32 \ion{H}{i} absorbers at 0.2 $<$ $z$ $<$ 1.4 and 79 associated galaxies, with available or upcoming molecular gas measurements from ALMA. 
We use a dedicated 40-orbit HST UVIS and IR WFC3 broad-band imaging campaign to characterise the stellar content of these galaxies. By fitting their spectral energy distribution, we establish they probe a wide range of stellar masses: 8.1 $<$ log($M_*$/M$_{\odot}$) $<$ 12.4. 
Given their star-formation rates, most of these objects lie on the main sequence of galaxies.
We also confirm a previously reported anti-correlation between the stellar masses and CGM hydrogen column density N(\ion{H}{i}), indicating an evolutionary trend where higher mass galaxies are less likely to host large amounts of \ion{H}{i} gas in their immediate vicinity up to 120 kpc.
Together with other studies from the MUSE-ALMA Haloes survey, these data provide stellar masses of absorber hosts, a key component of galaxy formation and evolution, and observational constraints on the relation between galaxies and their surrounding medium.
\end{abstract}

\begin{keywords}
galaxies: evolution -- galaxies: stellar content --
quasars: absorption lines
\end{keywords}



\section{Introduction}

One of the key questions in galaxy evolution is how galaxies interact and connect to their immediate environment, called the circumgalactic medium (CGM, \citealp{Tumlinson2017}). 
Once stars are formed, galaxies expel ionising photons and heavy elements formed in stars and supernovae into their surrounding environment through galactic winds \citep{Pettini2008,Shull2014}. 
Any gas flows into and out of the galaxy traverse through this extended gas halo and therefore the CGM is closely connected to the galaxy's evolution \citep{Muratov2015,Fox2017}. 
A detailed investigation of gas inflows and outflows is of paramount importance for understanding these processes. 
Since gas, stars, and metals are intimately connected, gas flows affect the history of star formation and chemical enrichment in galaxies. 
Therefore the study of the multi-phase CGM (extending over hundreds of kpc around galaxies - \citealt{Shull2014}) is crucial for understanding the star formation and galaxy evolution as a whole.

However, determining what drives the physical processes at play in the CGM still remains a complex problem in galaxy evolution, in large part due to the lack of significant observational constraints. 
Direct observations of this gas halo are challenging as the gas is extended and diffuse and the expected emission from the CGM is faint and mostly below current instrument detection limits \citep{Augustin2019,Corlies2020}.

In order to effectively explore the diffuse gas in the Universe, absorption-line spectroscopy proves to be a valuable method, as its detection sensitivity remains unaffected by redshift. 
One particularly compelling approach is to detect absorption lines in the spectra of bright background quasars, which allows for a comprehensive study of the distribution, chemical properties, and kinematics of CGM gas (e.g. \citealt{Prochaska2005,Tumlinson2013}).

Typically, strong (log [$N$(H I)/cm$^{-2}$]$\geq$18) absorbers observed in the lines-of-sight to high redshift quasars serve as suitable targets to be followed up with additional imaging and spectroscopy to determine its relation to galaxies. 
A comprehensive understanding of the baryon cycle can be achieved by combining quantitative data on stellar properties with the already established gaseous measurements.
A significant number of studies have successfully identified hosts to so-called DLAs (Damped Lyman-$\rm \alpha$ systems) and sub-DLAs \citep{Peroux2003}, which are the strongest \ion{H}{i} absorbers \citep{Augustin2018,Rhodin2018,Krogager2017}.
Due to their high column density these absorbers host a significant amount of neutral gas \citep{PerouxHowk} and are therefore believed to be closely connected to their host galaxies. 
This makes them ideal targets to study the relation between galaxies and their CGM.

Recently, 3D Integral Field Spectroscopy (IFS), which produces data cubes where each pixel on the image has a spectrum, has provided a novel technique to examine the gas in absorption against background sources whose lines of sight pass through a galaxy's CGM. 
Building on early efforts with Near-Infrared (NIR) Integral Field Spectrograph (IFS) SINFONI (Spectrograph for INtegral Field Observations in the Near Infrared) on the Very Large Telescope (VLT) \citep{Bouche2007, Peroux2011, Peroux2013, Peroux2016,Augustin2018}, the potential of this technique for studying the CGM with the optical IFU VLT/MUSE (Multi Unit Spectroscopic Explorer) has been demonstrated \citep{Schroetter2016, Bouche2016, Zabl2019, Fumagalli2016, Muzahid2020, Dutta2021, Berg2022}. 
Using this technique, a novel survey has been designed over the past years, combining IFU data from VLT/MUSE with ALMA and HST data of 19 quasar fields, labeled MUSE-ALMA haloes \citep{Hamanowicz2020,Peroux2022, Weng2023, Karki2023}.
Within these fields, 79 galaxies were identified to be associated within 500 km/s of strong quasar absorbers at $z\sim$ 1 \citep{Peroux2022}. 
By probing the neutral, molecular and ionised gas around these galaxies, the physical properties of the CGM have been characterised from the combination of abundance determination and kinematics through Voigt profile fitting of the absorber and emission-line and stellar continuum luminosities of the host galaxies \citep{Augustin2018, Klitsch2018, Peroux2019, Hamanowicz2020}.

The goal of the present work is to compute the stellar mass of galaxies related to gas-rich absorbers and understand what population of galaxies is traced by those absorbers. 
Particularly, we want to revisit the previously identified stellar mass - \ion{H}{i} column density anticorrelation \citep{Augustin2018} and fit its slope.
Additionally, we determine whether the galaxies that are found as absorber hosts have physical properties which differ from the general population of galaxies, whether they are more star forming or quenched than the general galaxy population or whether they are following the same scaling relations as the general galaxy population.
Particularly, \cite{Peroux2020} have shown that the metal distribution around galaxies is anisotropic. 
Their findings indicate that the metallicity trend with azimuthal angle is strongly dependent on the stellar mass of the host, indicating that the stellar mass is a critical component to the structure and evolution of galactic haloes.
In this work we determine the stellar masses of galaxies associated with absorbers, providing essential information on these galaxies to better understand their relation to the gas seen in absorption.

The manuscript is organized as follows: 
Section \ref{sec:obs} presents the observations used in this study. 
Section \ref{sec:sed} details the process of fitting Spectral Energy Distribution to the data, while 
Section \ref{sec:prop} relates the stellar properties of these galaxies with their CGM characteristics. 
We summarize and conclude in Section \ref{sec:conclusions}. 
Here, we adopt an $H_0$ = 67.74 km s$^{-1}$ Mpc$^{-1}$, $\Omega_M$ = 0.3089, and $\Omega_{\Lambda}$ = 0.6911 cosmology.

\section{MUSE-ALMA Haloes Observations} \label{sec:obs}

\subsection{Survey overview and relevant data}

Integral Field Units (IFUs) have opened a new era in establishing the relation between absorption and emission. 
The optical IFU VLT/MUSE \citep{Bacon2010} has proven to be a true game-changer in the field. 
The MUSE-ALMA Haloes survey probes the multi-phase CGM gas of intermediate redshift galaxies. 
The main goal of the survey is to reveal and understand the physical processes responsible for the transformation of baryons in galaxies \citep{Peroux2022}. 
The survey is based on a unique selection of known quasar absorbers with measured \ion{H}{i} column density log [$N$(H I)/cm$^{-2}$]$\geq$18, from HST UV spectroscopy with resolutions of $R$=20,000--30,000. 
\cite{Weng2023} have measured metal lines from optical spectroscopy and analysed the physical and emission-line properties of 79 galaxies associated with 32 \ion{H}{i} absorbers at redshift $0.20 \lesssim z \lesssim 1.4$. 
These associated galaxies are selected to have velocities $\leq 500$ $\rm km s^{-1}$ relative to the absorber redshift. They are found at impact parameters
 of $\sim$5.7-100 kpc at the lower redshifts and up to 270 kpc at the higher redshifts, where the upper cut-off reflects the field-of-view of MUSE. 
The star-formation rates (SFRs) of associated galaxies are measured using the H$\alpha$ emission line when available. For sources at redshift $z$ $\geq$ 0.4, where H$\alpha$ is not observable with MUSE, \cite{Weng2023} estimate the SFR using the [\ion{O}{ii}] luminosity. For galaxies with dust corrections available, they also provide a dust-corrected SFR. 
3-$\sigma$ SFR limits are calculated for non-detections. In the resulting sample, there are both passive galaxies without detectable emission lines, and star-forming galaxies with SFRs up to 15 $\rm M_{\odot} yr^{-1}$.

\subsection{HST Broad-band Imaging Data}

MUSE-ALMA Haloes also includes broad-band imaging of all but one field in the sample. This includes HST WFC3 imaging data in selected near-UV and optical filters, combined with archival Ultra-Violet, optical and Infra-Red WFPC2 and WFC3 imaging data to investigate the immediate surroundings of the quasar absorbers \citep{Peroux2022}. The filters for the observations are carefully designed to surround the 4000 \AA\ break at each absorber's redshift.
For the majority (16/19) of the fields images in three broad-band filters are available. 
Using additional data available in the HST archive, we cover a total of four filters for some of the targets. 
The MUSE IFU observations of the quasar fields are used to design the HST observations. 
We identified [OII] emission from galaxies at the redshift of the known absorbers as potential host galaxies and determined their location with respect to the quasar.
The locations of these galaxies are used to optimize the HST pointings such that the diffraction spikes from the HST optics would not spatially overlap with the galaxies, while simultaneously keeping the whole MUSE FOV within the HST field of view.
We then performed a careful quasar PSF subtraction, following \cite{Augustin2018}, and photometry measurements (as described in detail in  \citealt{Peroux2022,Karki2023}).
The \textsc{Astropy} package \textsc{Photutils} is used on the processed HST broad-band images in each filter for each field to search for all objects and to perform photometry of the detected objects.
The details are described in \citet{Karki2023}.
In cases where an object detected in one filter was not detected in another filter, a 3-$\sigma$ magnitude limit was calculated for the filter with the non-detection by measuring the 3-$\sigma$ noise level within the same aperture as for the detection.
The resulting magnitudes are publicly released with \citet{Peroux2022} and are used for the SED fitting described below.

\section{Spectral Energy Distribution fitting} \label{sec:sed}

Building on the successes of earlier multi-band approaches (e.g. \citealt{Christensen2014, Augustin2018}), we perform Spectral Energy Distribution (SED) fits of the galaxies associated with the absorbers.
The wavelength coverage and SNR of the continuum emission of the galaxies in question in the VLT/MUSE observations alone does not provide sufficient data for state-of-the-art SED algorithms to converge for the redshifts we are probing. 
Instead, the new HST data cover wavelengths around the 4000 \AA\ break and are designed to detect the stellar continuum of absorber counterparts in order to probe the stellar population of these galaxies. 
The coverage of this break significantly improves the accuracy of the stellar mass estimates, so we ensure that we obtain observations on both sides of the break. 
Additionally, the strength of the break directly affects the ability to measure the stellar masses effectively.
We note that, highly star-forming galaxies have a less defined break than quenched galaxies. 
Therefore the masses of quenched galaxies are determined more robustly by the fitting procedure.

We use the \textsc{Le Phare} \citep{Arnouts1999,Ilbert2006} SED fitting code to determine the physical properties of the observed galaxy sample. 
We assume the \cite{Bruzual2003} SED library, a \cite{Calzetti1994} dust extinction law and a \cite{Chabrier2003} initial mass function. 
We focus here on the galaxies whose redshifts have been measured with VLT/MUSE spectroscopy and are known to be within 500 $\rm km s^{-1}$ from the strong \ion{H}{i} absorbers. 
The resulting sample comprises 79 galaxies for 31 absorbers {(see Table \ref{tab:masses})}. 
To perform the SED fitting, we use the reliably determined magnitudes based on the high spatial resolution HST imaging. 
Some of the faint objects are detected in VLT/MUSE through their line-emission signatures, but not in continuum in the broad-band HST images, particularly in the bluest filters. {In cases where galaxies had no detection in any HST filter, we make use of the photometry catalogs from the Dark Energy Legacy Survey (DECaLS, \citealt{Dey2019}) to perform the SED fitting. These systems are labeled as "LEGACY" in the last column of table \ref{tab:masses}.
For galaxies with detections in one of the three bands, the SED fitting provides less robust stellar mass estimates, dubbed "probable".} {For illustration, we show in the Appendix examples of SED fits for a robust and a probable system (Fig. \ref{fig:SED_examples}).} These estimates are included in our sample for completeness, 
while we highlight the more reliable stellar mass measurements. 
We therefore focus here on the 34 galaxies with the most robust fits.

\begin{figure}
    \centering
    \includegraphics[width=.5\textwidth]{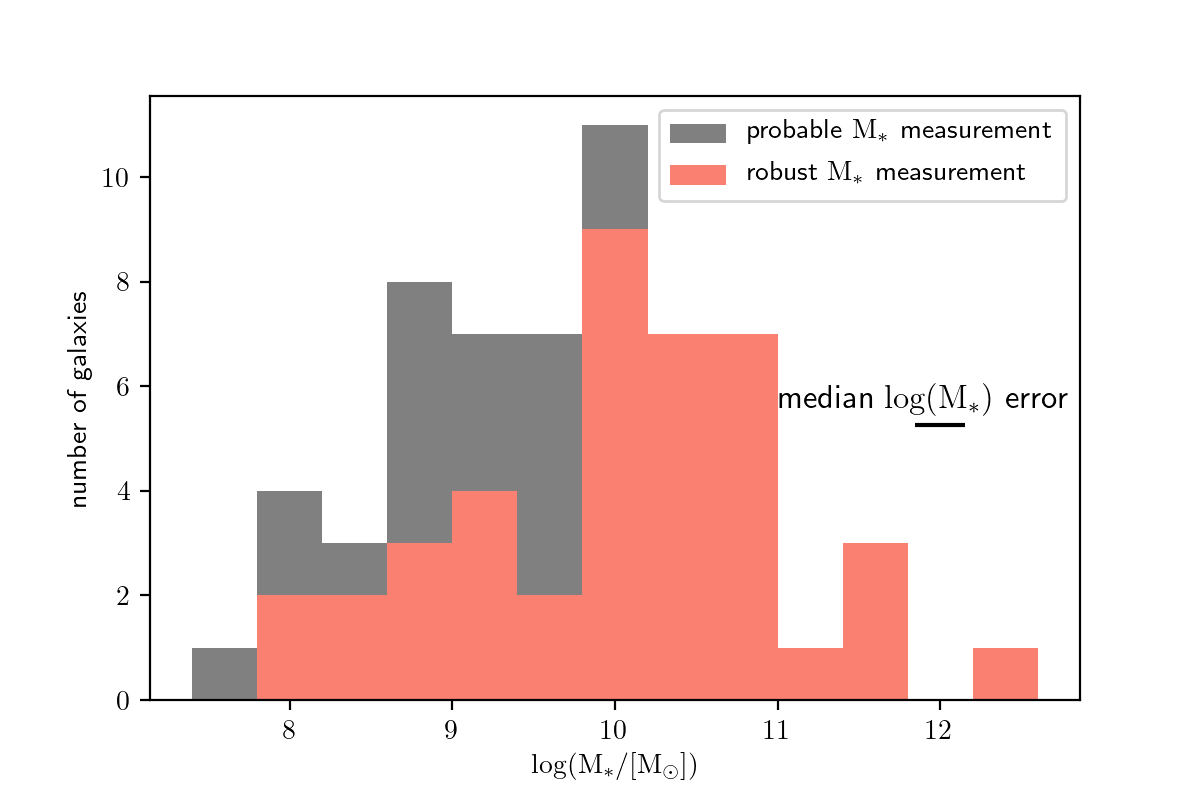}
    \caption{Distribution of stellar masses of absorber host galaxies: We find a large spread of stellar masses for absorber-associated galaxies, with log[$M_*$/M$_{\odot}$]= 7.8 -- 12.4.}
    \label{fig:robustmasses}
\end{figure}

Figure~\ref{fig:robustmasses} displays the distribution of stellar masses resulting from the SED fits. 
The grey histogram indicates the possible $M_*$ determinations where fewer than two photometry measurements are available while the red histogram shows the more robust determinations where 2-4 photometry measurements are available.
The derived stellar masses cover a broad range from log [$M_*$/M$_{\odot}$]= 7.8 -- 12.4, which illustrates the diversity of these systems. 
We see that the galaxies with stellar masses above log[$M_*$/M$_{\odot}$]= 10 have their SED fit typically robustly determined while the lower mass galaxies often have non-detections in the bluer filters, limiting their stellar mass determination and potentially biasing our detection numbers and robust mass determinations towards higher masses.

We additionally run the \textsc{Le Phare} SED fitting code on all galaxies in the HST fields regardless of whether or not they have spectroscopic redshifts, in order to assess the potential of determining photometric redshifts with SED fitting.
We derive uncertainties of the order $\delta z \geq$0.5 in redshift, so we deem the results for such objects not robust enough to identify further absorber hosts beyond the ones detected through spectroscopy in MUSE cubes.

\section{stellar properties of absorber hosts} \label{sec:prop}

\begin{figure*}
    \centering
     \includegraphics[width=.49\textwidth]{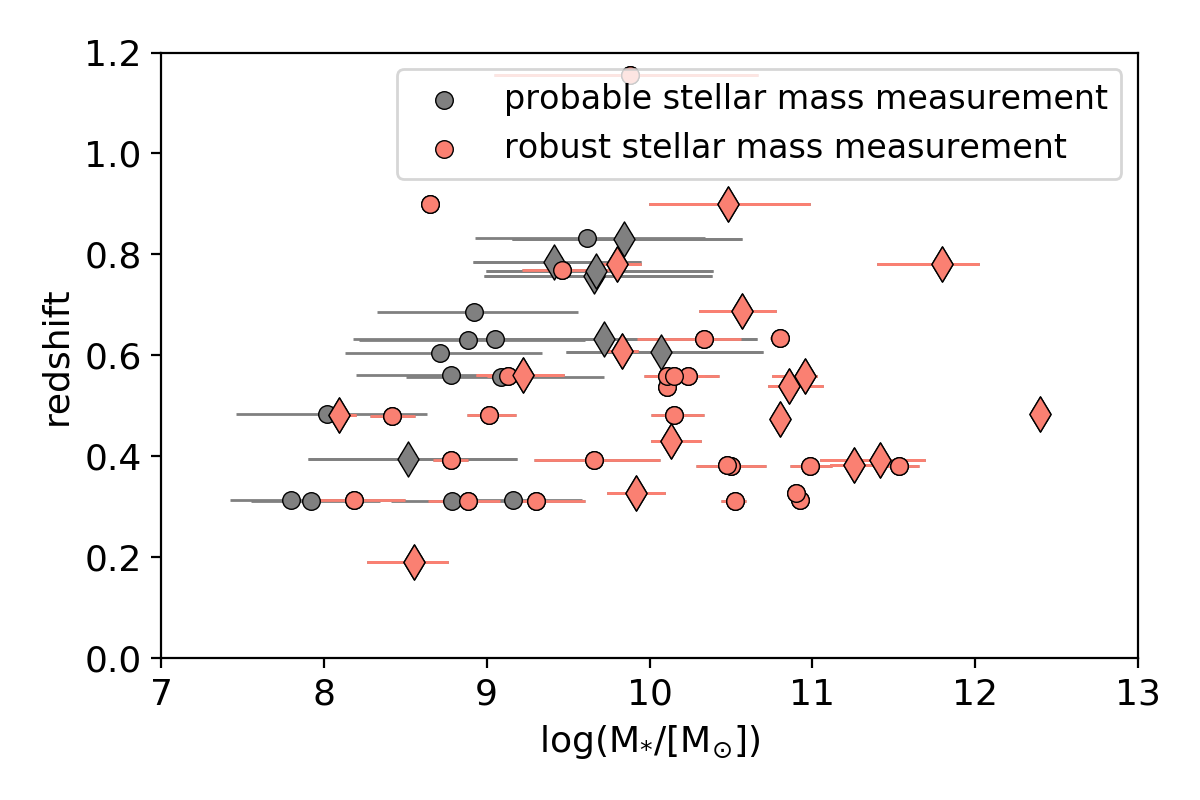}
    \includegraphics[width=.49\textwidth]{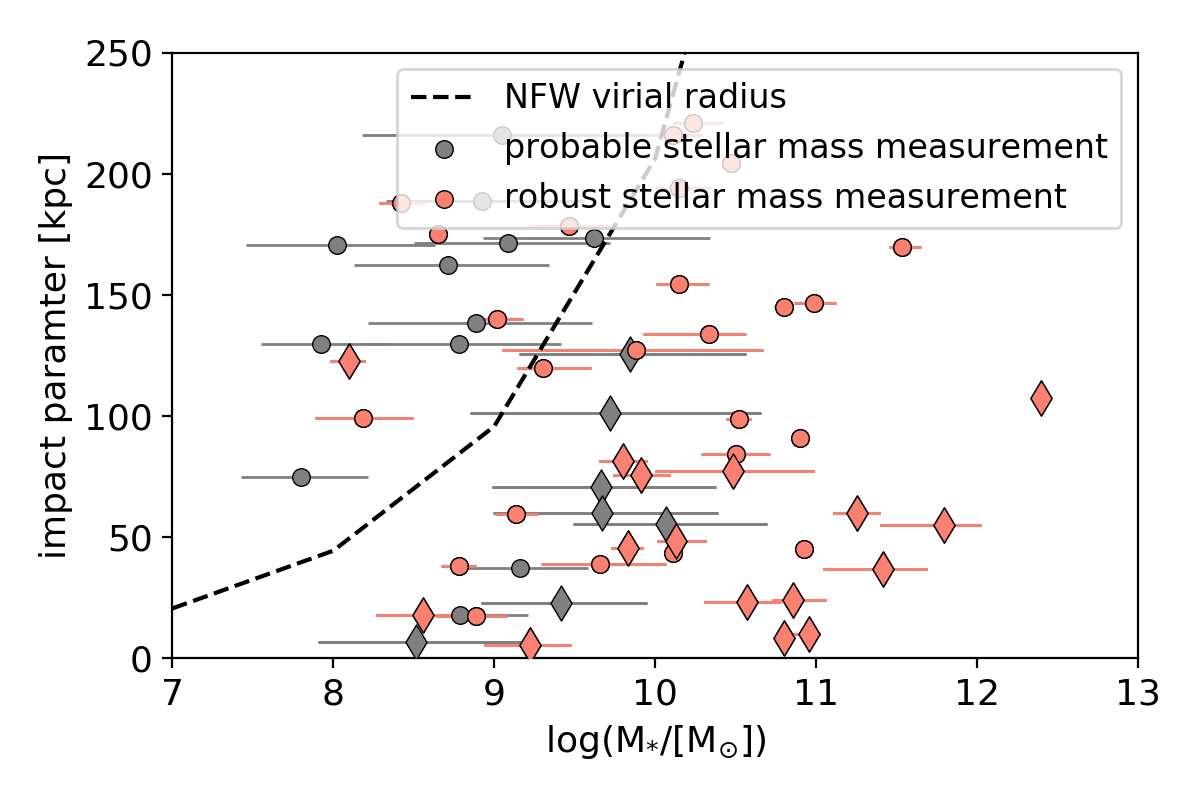}
    \includegraphics[width=.49\textwidth]{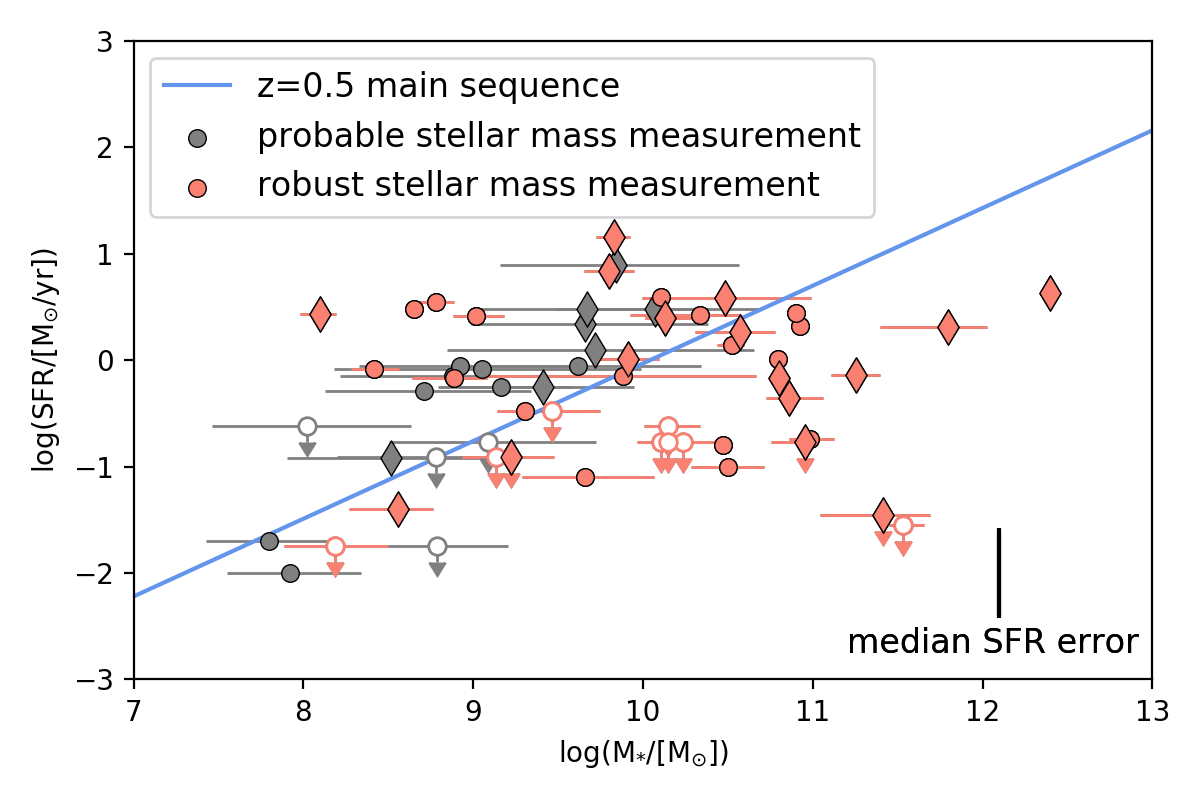}
    \includegraphics[width=.49\textwidth]{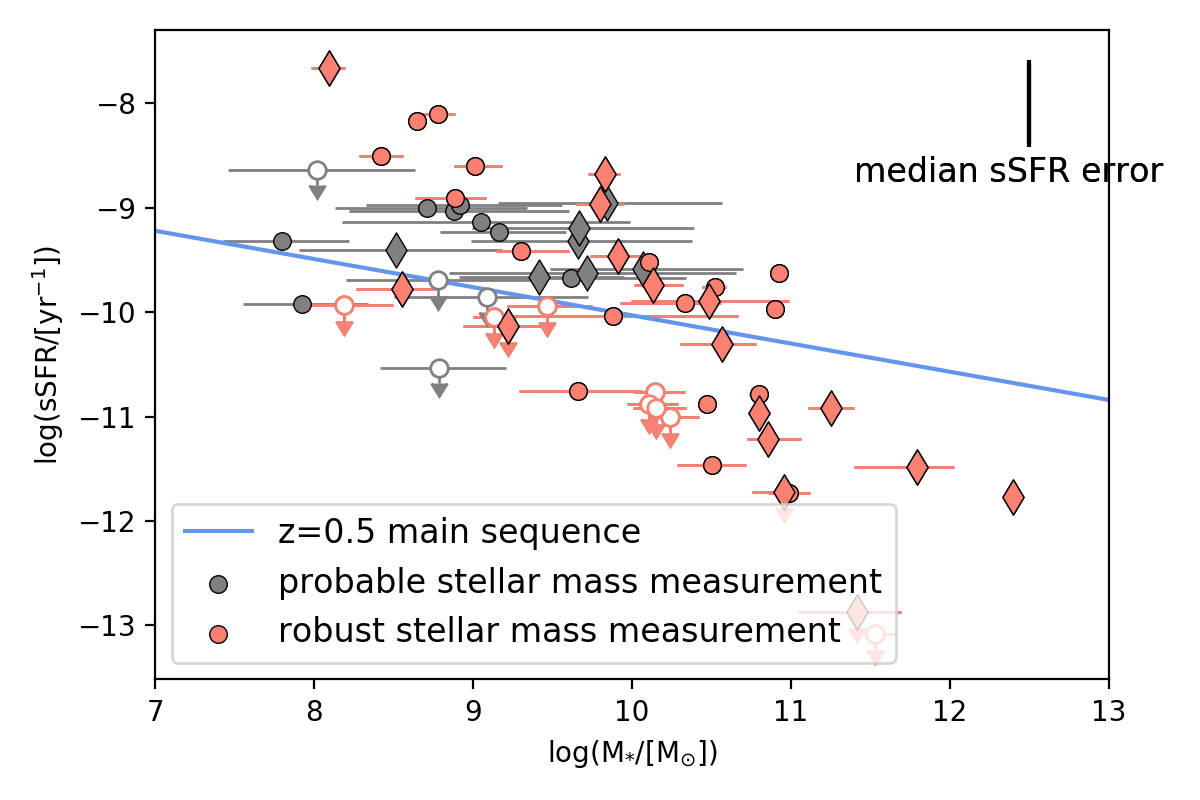}
    \caption{    Physical properties of the galaxies associated with gas-rich absorbers. In all these figures we denote robust stellar mass measurements in red and probable measurements in grey. The closest impact parameter galaxy of any system is marked with a diamond instead of a circle. 
    \textit{Upper left}: The redshift distribution of the determined stellar masses. We probe a range of stellar masses within 0.2$\leq$z$\leq$1.2, although there is a slight trend of finding the lower mass systems preferentially at lower redshifts. 
    \textit{Upper right}: The impact parameter range probed by the absorber hosts. We find no correlation between stellar masses and impact parameters. We consider the galaxy closest to the quasar sightline to be the main absorber host (diamond). The impact parameter range for these is up to $\sim$ 120 kpc.
    \textit{Bottom left}: We show how the absorber associated galaxies in our sample fall into the stellar mass - SFR plane in comparison to the main sequence at a matching redshift range \citep{Belfiore2018}. 
    We establish that most of these galaxies are normal star-forming galaxies and follow the main sequence. Some SFR measurements are found below the main sequence and could indicate quenching, particularly at the higher mass end.
    \textit{Bottom right}: Same as the Figure on the left, but showing the specific SFR (SFR over stellar mass) instead of absolute SFR, hightlighing the deviation of the higher mass galaxies from the star forming main sequence.
    }
    \label{fig:relations}
\end{figure*}
\begin{figure*}
    \centering
    \includegraphics[width=.49\textwidth]{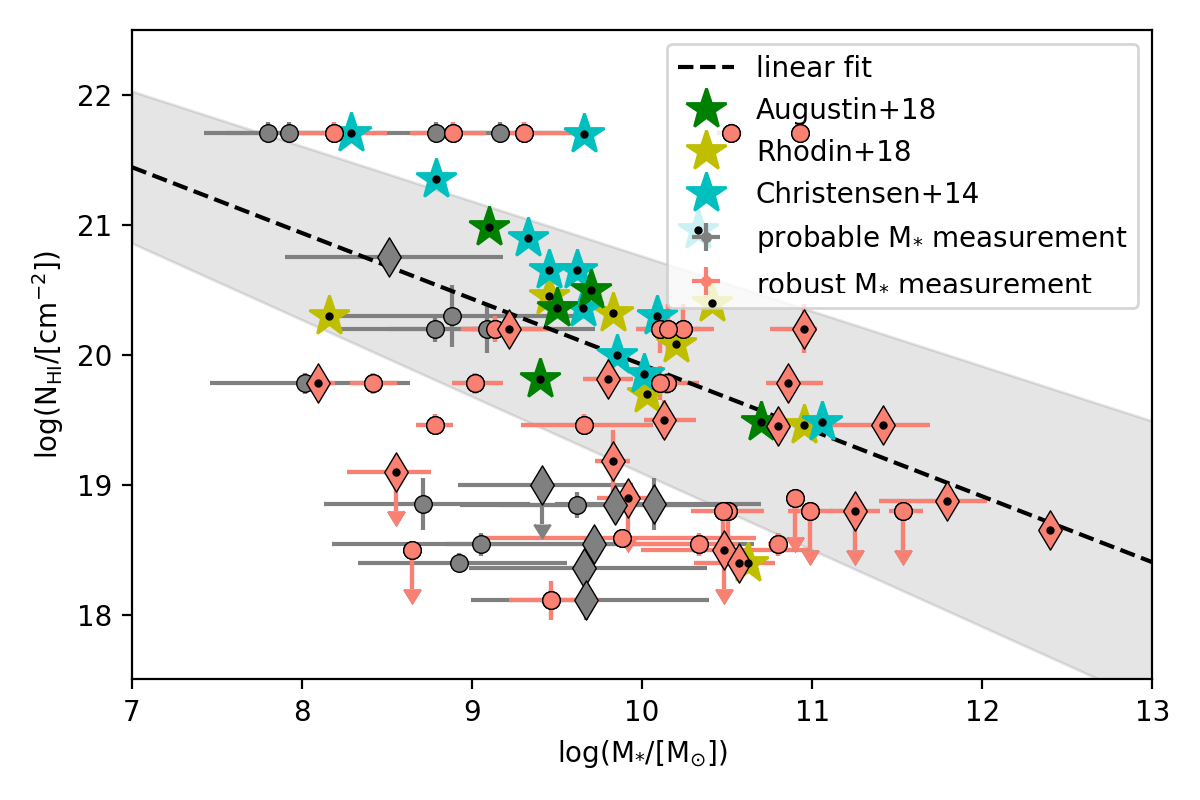}
    \includegraphics[width=.49\textwidth]{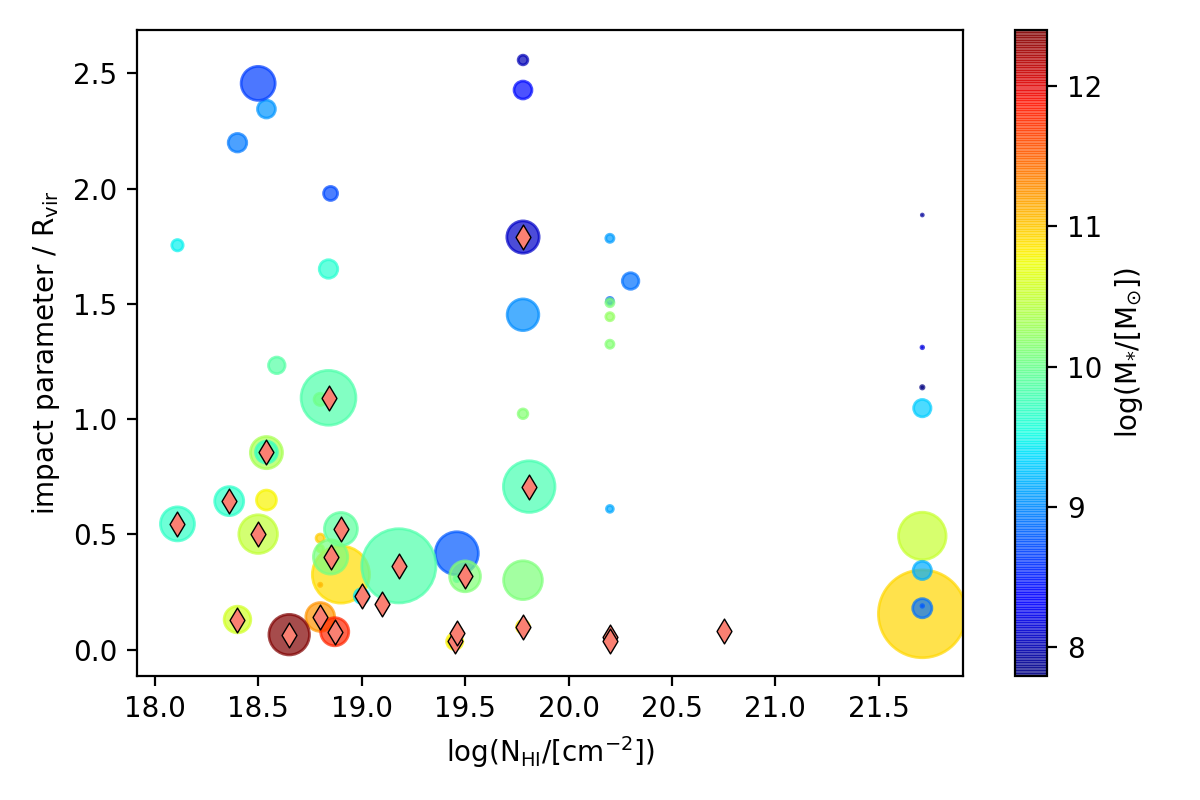}
    \caption{     
    \textit{Left}: The relation between \ion{H}{i} column density of the absorber versus the stellar masses of galaxies associated with those absorbers. 
    The literature data are taken from \citet{Augustin2018,Rhodin2018} and \citet{Christensen2014}.
    We find and quantify for the first time the anticorrelation between \ion{H}{i} absorption strength and stellar mass of the main host galaxy (dotted line). 
    The fit was performed on the stars and red diamond data points in the plot.
    \textit{Right}: Summary of all measured quantities in this work. 
    We show the {impact parameter scaled with virial radius of each system versus the neutral gas \ion{H}{i} column density}. The color of each circle indicates the stellar mass of that object and the size of the circle is proportional to the star formation rate of this galaxy. We find that the galaxies associated with high column density \ion{H}{i} absorbers span a range of SFR and stellar mass. As explored in the panel on the left, the highest column density systems seem to be hosted in groups of galaxies with smaller stellar masses, while the higher stellar mass objects are found at lower column densities. Simultaneously, at the higher column density end the host galaxies show relatively low SFRs compared to the galaxies found at the lower column density end of our sample.
    }
    \label{fig:relations2}
\end{figure*}

\subsection{Dependence of SFR, redshift and impact parameter on M$_*$}

First we investigate how the determined stellar masses depend on other properties of the galaxy that are connected to its evolution.

In the upper left panel of Figure \ref{fig:relations} we show the dependency of the determined stellar mass on the redshift of the galaxy.
Given that galaxies grow and evolve over time we expect to find generally more massive galaxies at lower redshifts and less massive galaxies at higher redshifts.
However, at the same time, due to surface brightness dimming, we expect to find preferentially bright and therefore massive galaxies at higher redshift and a range over all stellar masses at lower redshifts.
What we find in our sample is a mix of those effects. 
Overall we find a large scatter of stellar masses at all redshifts, however we do notice a lack of lower mass systems at redshifts higher than $z$ $\geq$ 0.7.

In the upper right panel of Figure \ref{fig:relations}, we investigate the dependency of the associated host stellar mass on the impact parameter to the absorber.
We find no correlation between these two quantities, indicating that galaxies of any mass can be found at any distance to a known absorber.
The diamonds indicate the sample of galaxies with lowest impact parameter for each absorption system, which are thought to be the main absorber hosts sample.
Here we see that the maximum impact parameter we probe with this sample of main absorber hosts is $\sim$120 kpc.
Following \cite{Behroozi2010} and \cite{Read2017}, we approximate the halo mass of the galaxies in our sample as $\rm M_{halo} \approx 100 \times M_{\star}$. The dashed line in Figure  \ref{fig:relations} shows the resulting virial radius, assuming a standard Navarro–Frenk–White (NFW) profile \citep{Navarro1996}.
We find that particularly at the lower mass end, a significant fraction of galaxies associated with the absorber lie outside the virial radius. 
{As in any absorber-host-connection study, there is a possibility that the closest galaxy contributing most the absorbing gas lies directly close to the projected position of the background quasar and/or is too faint to be detected. \cite{Weng2024} have recently quantifies this effect using hydrodynamical cosmological TNG50 simulations. Therefore, we cannot fully exclude the possibility of missing faint galaxies close to the quasar sightline.}

In the bottom row (left and right panels) of Figure \ref{fig:relations}, we show the measurements for stellar mass from SED fitting and SFR (left) and specific SFR (right) from emission lines in MUSE (H$\rm \alpha$ and [OII], \citealp{Weng2023}) compared to the expected main-sequence relation at $z$=0.5 \citep{Belfiore2018}.
We find that on average our sample of galaxies follows the main sequence, consistent with results of previous studies (e.g., \citealt{Kulkarni2022}).
However, at higher masses the galaxies fall slightly below the main sequence.
We recall that {a fraction (17 systems) of the SFR measurements is} dust corrected{, specifically those with H$\rm \alpha$ and H$\rm \beta$ measurements, corresponding to systems with $z$ $<$ 0.4 \citep[see][]{Weng2023}}. Therefore {estimates of SFR in the low-resdhift galaxies} represent a lower limit on the total SFR. 
{In particular, if the highest mass galaxies with log([$M_*$/[M$_{\odot}$])$>$ 11 were dust corrected, they might then well follow the main sequence.}
{We furthermore note that the combined sample of H$\rm \alpha$ and [OII] determined SFRs may introduce a larger scatter in the SFR distributions than a more homogeneous sample would.}
We note that the main sequence should also flatten at the higher mass end of the mass distribution \citep{Tomczak2016}.
Therefore we conclude that the sample of galaxies is representative of normal star-forming galaxies.
\citet{Langan2023} {and \citet{Guha2023}} find a similar result for \ion{Mg}{ii} absorption selected systems, indicating that they {follow the main sequence}.
Together with our findings, this indicates that absorber host galaxies follow the trends of the typical galaxies population.

\subsection{Correlation between M$_*$ and CGM atomic gas content}

One of the early findings of relating the \ion{H}{i} absorption properties with the host galaxy properties was an apparent anti-correlation between stellar mass of the host galaxy and the absorption \ion{H}{i} column density \citep{Augustin2018}.
Such an anti-correlation seems counter-intuititive and may be caused by additional factors such as measuring the impact parameter between the galaxy and the absorber, star formation rate of the host, or the environment of the galaxy. 
While previous studies were characterising somewhat limited sample sizes, we make use here of the currently complete MUSE-ALMA Haloes survey sample, covering quasar fields with 32 known absorbers and 79 associated galaxies with those absorbers, out of which we have robust masses for 34 galaxies. 
We are therefore testing this relation over a larger redshift and column density range and in more environments than previously possible.

In the left panel of Figure \ref{fig:relations2} we show for our sample the relation between the stellar mass of the host and the detected \ion{H}{i} column density in its CGM.
While previous studies that found this anti-correlation \citep{Augustin2018,Rhodin2018} considered objects where each absorber was associated with only one - typically the closest - galaxy, here we are considering more complex systems, including groups of galaxies associated with a strong \ion{H}{i} absorber.
The reason for the single detections in the literature data, particularly the ones using SINFONI \citep{Augustin2018} is the somewhat limited field-of-view around the quasar.
To simplify and homogenize this investigation and make it comparable to previous studies we consider here the galaxy at the closest impact parameter, marked with a diamond, to be the main absorber host galaxy.
The reason behind choosing the closest galaxy in impact parameter is the assumption that the physically closest galaxy to the gas we are probing in absorption will have the largest contribution to the absorption signal of all the galaxies within a given group \citep{Weng24}.

Now, we investigate this previously identified anti-correlation between stellar mass and circumgalactic \ion{H}{i} column density for the larger sample.
We also add the data points from \citet{Christensen2014} for comparison and note that all of these previous DLA host galaxy studies, have a larger average and greater spread in redshift, whereas our sample is at a slightly lower redshift on average.

Assuming an anti-correlation power law between stellar mass and \ion{H}{i} column density, we fit the relation. {We note that the upper limits were not considered in the fit. The data points used for the fit are marked with black dots in the left panel of Figure \ref{fig:relations2}. } 
For the fit we consider robust $M_{*}$ measurements of the closest impact parameter galaxies (red data points with diamond) as well as the literature samples from \cite{Christensen2014}, \cite{Augustin2018} and \cite{Rhodin2018}.
These literature samples consist of single counterparts detections of DLAs and subDLAs and showed a tentative trend for this anti-correlation, which we now have enough data points to actually measure.
Our choice of considering the closest galaxy in line of sight to the quasar homogenizes the chosen sample for the fit of the anticorrelation.
{We perform a Pearson correlation test and find a value of $-$0.57 thus confirming the anticorrelation. Interestingly, we note that the trend is driven by the literature subsample. When focussing on the new measurements from this work, we find a Pearson coefficient of $-$0.36, indicating little anticorrelation.}
We fit a simple power-law to the relation in logarithmic space:

\begin{equation}
   \rm  log(\textit{N}_{HI}/[cm^{-2}]) = -0.51 (\pm 0.1) \times log(\textit{M}_{*}/[M_{\odot}]) + 25.0 (\pm 1.2)
\end{equation}

The lack of high-mass high-$\rm \textit{N}_{HI}$ systems could be due to an observational bias or selection effect where high-$\rm \textit{N}_{HI}$, high stellar mass systems are artificially excluded. 
Indeed, a high-$\rm \textit{N}_{HI}$, high stellar mass system could arise in the center of a massive, potentially dusty galaxy, causing a reddening of the background quasar, which we would miss from typical quasar optical selection \citep{Vladilo2005}.

However, assuming this anti-correlation, or rather, upper envelope in detected $\rm \textit{N}_{HI}$ - stellar mass values has a physical origin, it holds information on the composition and evolution of galactic haloes. The lack of high-mass high-$\rm \textit{N}_{HI}$ systems can intuitively be explained by the ambient temperature of higher mass galaxies being typically higher than in lower mass galaxies, due to stronger feedback heating up the halo (see e.g. \citealt{Suresh2017}). 
This hotter gas may prevent the survival of cooler gas clouds hosting HI, at least within the inner $\sim$ 120 kpc as probed by our sample of closest impact parameter hosts.
Likewise, the higher mass galaxies have undergone a rapid star forming process and already used up their cool \ion{H}{i} gas reservoir, whereas lower mass galaxies take longer to use their gas supplies and still maintain more of it in their CGM.
As we probe lower column density systems, the spread in stellar mass of associated galaxies increases, with the mean shifting towards higher masses.
This means that we can find lower column density systems around a variety of galaxies but preferentially around higher mass systems as compared to the high $\rm \textit{N}_{HI}$ absorbers.
In a lower column density regime, \cite{Berg2022} find a scatter of stellar masses log([$M_*$/[M$_{\odot}$])$\sim$ 8.5 -- 12  for absorber systems with $N_{\ion{H}{i}}$ $<$ $10^{17}$ $\rm cm^{-2}$  without any clear trend. Similarly \cite{Chen2019} find column densities of $\sim$ $10^{18}$ $\rm cm^{-2}$ around galaxies with log([$M_*$/[M$_{\odot}$])$\sim$ 7.5 -- 11.5.
A reason for this phenomenon could be that more massive galaxies reside in larger haloes, providing a larger cross section for low column density absorption systems than lower mass galaxies.
Our results are also in line with \citet{Khare2007} and \citet{Kulkarni2010} who argue that sub-DLAs hosts are more massive than DLA hosts.
Finally, {\citet{Dutta2023} recently found that Ly$\rm \alpha$ rest-frame equivalent widths show a peak at log([$M_*$/[M$_{\odot}$])$\sim$ 9 and decline at both high and low masses. They attributed this trend to high virial temperature and efficient feedback in high mass halos. Since most of the galaxies in the present study have log([$M_*$/[M$_{\odot}$])$>$ 9, and while we considered column densities rather than equivalent widths, our observed anti-correlation is aligned with their results.}

Taken together, these findings suggest an evolutionary trend of the CGM composition with stellar mass where lower mass systems host haloes that are abundant with cool and dense \ion{H}{i} gas and higher mass systems are more depleted of cool gas.
However, even the more massive galaxies are fairly normally star forming, given the depletion of fuel in their immediate surroundings. 
More detailed studies of the gas content around galaxies and the galaxy properties such as stellar mass and star-formation rate but also metallicity, can measure the accretion timescales. 
These timescales describe the gas flows from the halo onto the disk and their transformation into stars versus the timescales for heating up the halo through feedback from a growing galaxy and depleting the gas fuel.

\section{Conclusions} \label{sec:conclusions}

In this work, we have performed stellar mass measurements of 79 0.2$<$z$<$1.4 galaxies known to have velocities within 500  $\rm km s^{-1}$ from a strong N(\ion{H}{i}) absorber. To this end, we have used \textsc{Le Phare} SED fitting on multi-broad-band HST imaging of the fields to robustly determine the stellar masses of the subset of 34 galaxies that were detected in at least two imaging bands. 

Our main results are:

\begin{itemize}
    \item We find that galaxies associated with high \ion{H}{i} column density absorbers span a large range in stellar masses of log([$M_*$/[M$_{\odot}$])= 8.1 -- 12.4 at 0.2$<$z$<$1.4. There is no siginificant trend with redshift or impact parameter. All these galaxies are normal star-forming galaxies following the expected main-sequence relation.

    \item A previously tentative detection of an anti-correlation between the stellar mass and CGM \ion{H}{i} column density has been confirmed by the larger sample of absorber-host pairs, indicating an evolutionary trend of cool gas depletion in the CGM with stellar mass.
\end{itemize}

Ultimately, connecting the CGM gas properties of galaxies with their stellar content provides fresh clues on the baryon cycle, a key component of galaxy formation and evolution.
In particular, knowledge of the stellar masses of galaxies allows to perform studies of the azimuthal metallicity distribution \citep{Weng2023b}.
Further studies of the detailed chemical composition of absorber host galaxies may reveal the timescales involved in gas accretion and recycling.
Cosmological simulations are now challenged to resolve both the cool gas in the CGM as well as the properties of the host galaxies in order to provide physical ground to our findings and make predictions to the galactic baryon cycle that can be tested by such observations.

\section*{Acknowledgements}

The authors would like to thank Dylan Nelson, Max Pettini and Jason Tumlinson for useful discussions and constructive feedback on this work.
This research was supported by the International Space Science Institute (ISSI, \url{https://www.issibern.ch/}) in Bern, through ISSI International Team project \#564 (The Cosmic Baryon Cycle from Space). 
RA acknowledges financial support from the STScI Director’s Discretionary Research Fund (DDRF) and funding by the European Research Council through ERC-AdG SPECMAP-CGM, GA 101020943. 
AK and VPK acknowledge support from a grant from the Space Telescope Science Institute for GO program 15939 (PI: P{\'e}roux), and additional partial support from US National Science Foundation grant AST/2007538 and NASA grant  80NSSC20K0887 (PI: Kulkarni). 
SW acknowledge the financial support of the Australian Research Council through grant CE170100013 (ASTRO3D). 
GGK acknowledges the support of the Australian Research Council through the Discovery Project DP170103470.
A.K.~gratefully acknowledges support from the Independent Research Fund Denmark via grant number DFF 8021-00130. 

In this work we used the following python packages:
Numpy \citep{harris2020array}; Matplotlib \citep{Hunter:2007}; pandas \citep{mckinney-proc-scipy-2010,reback2020pandas}; AstroPy \citep{astropy}; Halotools \citep{Hearin2017}
\section*{Data Availability}

All data presented in this work is publicly available or available upon request.



\bibliographystyle{mnras}
\bibliography{main} 



\begin{table*}
 \begin{tabular}{lrll|lrll}
\hline \hline
             ID &  Filters &                  Mass &   Notes &              ID &  Filters &                  Mass &   Notes \\
             \hline
 Q0138m0005\_14 	&	3	&	   9.8$_{-0.2}^{+0.1}$ 	&	     HST 	&	  Q1130m1449\_68 	&	1	&	   7.8$_{-0.4}^{+0.4}$ 		     & HST \\
  Q0152m2001\_4 	&	2	&	  11.5$_{-0.1}^{+0.1}$ 	&	     HST 	&	  Q1130m1449\_76 	&	0	&	                    N/A 		     & HST \\
  Q0152m2001\_5 	&	2	&	  11.3$_{-0.1}^{+0.1}$ 	&	     HST 	&	   Q1211p1030\_7 	&	3	&	   8.8$_{-0.1}^{+0.1}$ 		     & HST \\
  Q0152m2001\_7 	&	2	&	  11.0$_{-0.1}^{+0.1}$ 	&	     HST 	&	   Q1211p1030\_9 	&	2	&	   9.7$_{-0.4}^{+0.4}$ 		     & HST \\
 Q0152m2001\_12 	&	2	&	  11.8$_{-0.4}^{+0.2}$ 	&	     HST 	&	  Q1211p1030\_13 	&	3	&	   8.7$_{-0.1}^{+0.1}$ 		  & LEGACY \\
 Q0152m2001\_13 	&	2	&	  10.5$_{-0.2}^{+0.2}$ 	&	     HST 	&	  Q1211p1030\_16 	&	2	&	  10.5$_{-0.5}^{+0.5}$ 		     & HST \\
 Q0152m2001\_14 	&	3	&	  10.5$_{-0.1}^{+0.1}$ 	&	  LEGACY 	&	  Q1211p1030\_17 	&	3	&	  11.4$_{-0.4}^{+0.3}$ 		  & LEGACY \\
 Q0152m2001\_62 	&	0	&	                    N/A 	&	     HST 	&	  Q1211p1030\_38 	&	1	&	   8.9$_{-0.7}^{+0.7}$ 		     & HST \\
  Q0152p0023\_7 	&	2	&	  10.1$_{-0.1}^{+0.2}$ 	&	     HST 	&	  Q1211p1030\_48 	&	0	&	                    N/A 		     & HST \\
 Q0152p0023\_13 	&	2	&	   9.0$_{-0.1}^{+0.2}$ 	&	     HST 	&	  Q1211p1030\_57 	&	0	&	                    N/A 		     & HST \\
 Q0152p0023\_20 	&	2	&	   8.1$_{-0.1}^{+0.1}$ 	&	     HST 	&	  Q1211p1030\_58 	&	0	&	                    N/A 		     & HST \\
 Q0152p0023\_23 	&	2	&	   8.4$_{-0.1}^{+0.1}$ 	&	     HST 	&	    Q1229m021\_5 	&	3	&	   9.5$_{-0.2}^{+0.3}$ 		  & LEGACY \\
 Q0152p0023\_44 	&	1	&	   8.0$_{-0.6}^{+0.6}$ 	&	     HST 	&	    Q1229m021\_6 	&	0	&	                    N/A 		     & HST \\
  Q0420m0127\_8 	&	3	&	  10.3$_{-0.4}^{+0.2}$ 	&	  LEGACY 	&	    Q1229m021\_8 	&	1	&	   9.8$_{-0.7}^{+0.7}$ 		     & HST \\
 Q0420m0127\_12 	&	3	&	  10.8$_{-0.1}^{+0.1}$ 	&	  LEGACY 	&	   Q1229m021\_10 	&	1	&	   9.6$_{-0.7}^{+0.7}$ 		     & HST \\
 Q0420m0127\_13 	&	1	&	   9.7$_{-0.9}^{+0.9}$ 	&	     HST 	&	   Q1229m021\_13 	&	1	&	   9.7$_{-0.7}^{+0.7}$ 		     & HST \\
 Q0420m0127\_30 	&	1	&	   9.1$_{-0.9}^{+0.9}$ 	&	     HST 	&	   Q1229m021\_29 	&	0	&	                    N/A 		     & HST \\
   Q0454m220\_4 	&	2	&	  12.4$_{-0.1}^{+0.1}$ 	&	     HST 	&	   Q1229m021\_39 	&	1	&	   8.5$_{-0.6}^{+0.7}$ 		     & HST \\
  Q0454m220\_69 	&	3	&	  10.8$_{-0.1}^{+0.1}$ 	&	  LEGACY 	&	   Q1229m021\_40 	&	1	&	   9.7$_{-0.7}^{+0.7}$ 		     & HST \\
  Q0454p039\_15 	&	0	&	                    N/A 	&	     HST 	&	  Q1229m021\_41 	&	0	&	                    N/A 		     & HST \\
  Q0454p039\_57 	&	0	&	                    N/A 	&	     HST 	&	   Q1342m0035\_4 	&	3	&	  10.1$_{-0.1}^{+0.1}$ 		     & HST \\
  Q0454p039\_65 	&	2	&	   9.9$_{-0.8}^{+0.8}$ 	&	     HST 	&	   Q1342m0035\_9 	&	2	&	  10.9$_{-0.1}^{+0.2}$ 		     & HST \\
  Q1110p0048\_6 	&	3	&	   9.1$_{-0.1}^{+0.1}$ 	&	     HST 	&	  Q1345m0023\_13 	&	1	&	  10.1$_{-0.6}^{+0.6}$ 		     & HST \\
 Q1110p0048\_15 	&	1	&	   8.8$_{-0.6}^{+0.6}$ 	&	     HST 	&	  Q1345m0023\_40 	&	1	&	   8.7$_{-0.6}^{+0.6}$ 		     & HST \\
 Q1110p0048\_44 	&	3	&	   9.2$_{-0.3}^{+0.3}$ 	&	     HST 	&	  Q1431m0050\_10 	&	3	&	   9.8$_{-0.1}^{+0.1}$ 		     & HST \\
  Q1130m1449\_4 	&	3	&	  10.9$_{-0.1}^{+0.1}$ 	&	     HST 	&	  Q1431m0050\_26 	&	2	&	  10.6$_{-0.3}^{+0.2}$ 		     & HST \\
  Q1130m1449\_5 	&	3	&	                    N/A 	&	     HST 	&	  Q1431m0050\_68 	&	0	&	                    N/A 		     & HST \\
  Q1130m1449\_6 	&	4	&	  10.9$_{-0.1}^{+0.1}$ 	&	     HST 	&	  Q1431m0050\_73 	&	1	&	   8.9$_{-0.6}^{+0.6}$ 		     & HST \\
  Q1130m1449\_8 	&	3	&	  10.5$_{-0.1}^{+0.1}$ 	&	     HST 	&	   Q1515p0410\_4 	&	3	&	  11.0$_{-0.2}^{+0.1}$ 		     & HST \\
  Q1130m1449\_9 	&	4	&	   9.9$_{-0.2}^{+0.2}$ 	&	     HST 	&	   Q1515p0410\_9 	&	2	&	  10.2$_{-0.1}^{+0.2}$ 		     & HST \\
 Q1130m1449\_13 	&	3	&	   9.3$_{-0.2}^{+0.3}$ 	&	     HST 	&	  Q1515p0410\_11 	&	2	&	  10.1$_{-0.1}^{+0.2}$ 		     & HST \\
 Q1130m1449\_16 	&	1	&	   9.2$_{-0.4}^{+0.4}$ 	&	     HST 	&	  Q1515p0410\_13 	&	2	&	  10.2$_{-0.1}^{+0.2}$ 		     & HST \\
 Q1130m1449\_17 	&	3	&	   8.9$_{-0.2}^{+0.2}$ 	&	     HST 	&	  Q1515p0410\_42 	&	1	&	   9.1$_{-0.6}^{+0.6}$ 		     & HST \\
 Q1130m1449\_18 	&	3	&	   8.6$_{-0.3}^{+0.2}$ 	&	     HST 	&	  Q1515p0410\_54 	&	0	&	                    N/A 		     & HST \\
 Q1130m1449\_24 	&	1	&	   8.8$_{-0.4}^{+0.4}$ 	&	     HST 	&	   Q1554m203\_51 	&	1	&	   9.4$_{-0.5}^{+0.5}$ 		     & HST \\
 Q1130m1449\_31 	&	0	&	                    N/A 	&	     HST 	&	   Q2131m1207\_5 	&	2	&	  10.1$_{-0.1}^{+0.2}$ 		     & HST \\
 Q1130m1449\_36 	&	0	&	                    N/A 	&	     HST 	&	  Q2131m1207\_26 	&	0	&	                    N/A 		     & HST \\
 Q1130m1449\_43 	&	0	&	                    N/A 	&	     HST 	&	  Q2131m1207\_34 	&	0	&	                    N/A 		     & HST \\
 Q1130m1449\_50 	&	2	&	   8.2$_{-0.3}^{+0.3}$ 	&	     HST 	&	  Q2131m1207\_43 	&	0	&	                    N/A 		     & HST \\
 Q1130m1449\_56 	&	1	&	   7.9$_{-0.4}^{+0.4}$ 	&	     HST 	&		&		&			      \\
\hline \hline
\end{tabular}

\caption{Stellar masses for 79 absorber host galaxies. The values are labelled "robust" if at least 2 detections in different HST filters were available for the SED fit. For the purpose of SED fit, photometry from the Dark Energy Camera Legacy Survey (DECaLS,https://www.legacysurvey.org/, \citealp{Dey2019}) is used for some objects. The "probable" values are based one photometry estimate or upper limits. These latter measurements are mainly present in the lower mass end of the distribution. }
    \label{tab:masses}
\end{table*}

\begin{appendix} 

\section{Spectral Energy Distribution Fit Resulting Spectra}

\begin{figure*}
    \centering
    \includegraphics[width=.49\textwidth]{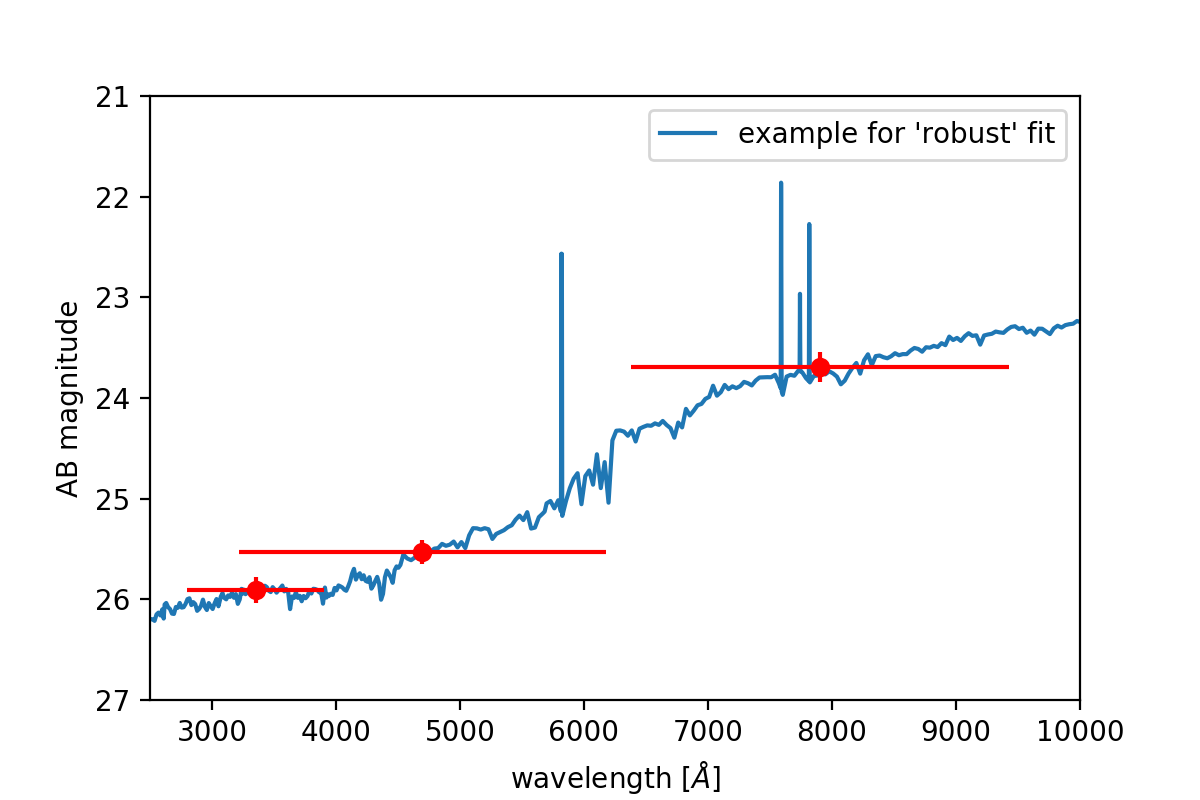}
    \includegraphics[width=.49\textwidth]{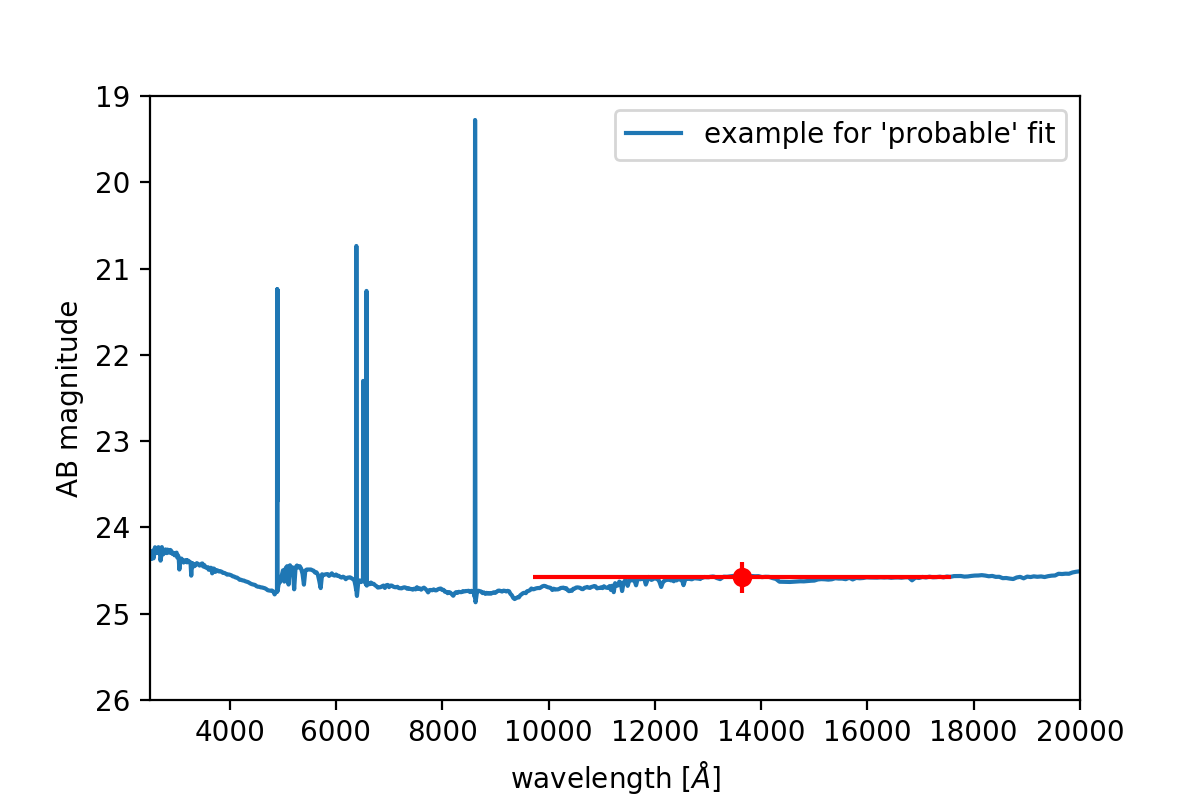}
    \caption{{The red points display the AB magnitudes measurements from the HST observations. The blue line represents the best fit galaxy spectrum derived from \textsc{Le Phare} algorithm. {\it Left:} An example of a fit based on detections of the galaxy in three HST bands and resulting in a fit providing a robust estimate of the stellar mass of the object. {\it Right:} An example of a fit based on a single magnitude measurement and providing a probable estimate of the stellar mass of that galaxy.}}
    \label{fig:SED_examples}
\end{figure*}

Figure~\ref{fig:SED_examples} displays two randomly chosen examples of Spectral Energy Distribution (SED) fits performed with \textsc{Le Phare} algorithm \citep{Arnouts1999,Ilbert2006}. 

\end{appendix} 


\bsp	
\label{lastpage}
\end{document}